# λ-transition in low dimensional systems with $SU_q(2)$ symmetry

Marcelo R. Ubriaco*

*Laboratory of Theoretical Physics*
*Department of Physics*
*University of Puerto Rico*
*P. O. Box 23343, Río Piedras*
*PR 00931-3343, USA*

## Abstract

We show that $SU_q(2)$ invariant systems in one and two dimensions exhibit Bose-Einstein condensation for $q > 1$. For these systems there is a λ-transition at the critical temperature. The critical temperature and the gap in the heat capacity increase more rapidly for small deviations from the standard value $q = 1$, and they become approximately constant for large values of $q$. For low temperatures and $q > 1$ the entropy is lower than the entropy of an ideal Bose gas.

---

*E-mail:ubriaco@ltp.upr.clu.edu

# 1 Introduction

More than seventy years have passed since the prediction that, in three dimensions and below a critical temperature $T_c$, the ground state of an ideal Bose gas becomes suddenly populated. This well known phenomenon, known as Bose-Einstein condensation (BEC) [1], and the conditions under which it occurs, has been the subject of theoretical research ever since.

Recent experimental realizations of BEC of $^7L_i$ [2], $^{87}R_b$ [3], and sodium [4] atoms have triggered a great deal of new theoretical work in this area. On the other hand, as it is a well known [5, 6], an ideal Bose gas does not exhibit BEC in one and two dimensions. The lack of a transition point is the consequence of the divergence of the function expressing the average number of particles above the ground state when the fugacity $z = 1$. Thus, in these cases the average energy $U$ and its temperature derivatives are smooth functions of the temperature $T$, and no interesting behavior occurs. Therefore, for many years a great deal of attention was paid to the study of one and two dimensional Bose systems interacting with an inhomogeneous external potential [5, 7, 8, 9, 10].

In this paper we study BEC for one and two dimensional systems described by the simplest quantum group [11] ($SU_q(2)$) invariant Hamiltonians with no external potential. These Hamiltonians are written in terms of a set of operators $\Phi_j$, $j = 1, 2$, whose algebra is covariant under the action of the quantum group $SU_q(2)$. Our main motivation is to learn whether imposing quantum group symmetries to a Bose gas has the effect of enhancing BEC, and, in that case, what is the behavior of the thermodynamic functions.

In Section 2 we briefly describe the Hamiltonian and the partition function of the model. For the interested reader, details of the formalism used in this paper can be found in References [14, 15]. In Section 3 we calculate the heat capacity $C_V$, critical temperature $T_C$ and entropy $S$ for our system in one and two dimensions. We show that, in contrast to the case of an ideal Bose gas, our model undergoes BEC with a critical temperature and a gap in the heat capacity, both increasing with the value of the parameter $q$. In addition, the entropy functions lie below the entropy of an ideal Bose gas.

## 2 Partition function

Our starting point is the simple Hamiltonian

$$\mathcal{H}_B = \sum_\kappa \varepsilon_\kappa (\mathcal{N}_{1,\kappa} + \mathcal{N}_{2,\kappa}), \tag{1}$$

with $[\overline{\Phi}_{i,\kappa}, \Phi_{\kappa',j}] = 0$ for $\kappa \neq \kappa'$ and $\mathcal{N}_{i,\kappa} = \overline{\Phi}_{i,\kappa} \Phi_{i,\kappa}$. The algebra satisfied by the fields $\Phi_j$ is given by the relations

$$\Phi_j \overline{\Phi}_i = \delta_{ij} + q R_{kijl} \overline{\Phi}_l \Phi_k \tag{2}$$

$$\Phi_l \Phi_k = q^{-1} R_{jikl} \Phi_j \Phi_i, \tag{3}$$

where the $R$-matrix

$$R = \begin{pmatrix} q & 0 & 0 & 0 \\ 0 & 1 & 0 & 0 \\ 0 & q - q^{-1} & 1 & 0 \\ 0 & 0 & 0 & q \end{pmatrix}.$$

Equations (2) and (3) are covariant under the action of the quantum group $SU_q(2)$, and they should not be confused with the $q$-boson algebra [12, 13]

$$a_i a_i^\dagger - q^{-1} a_i^\dagger a_i = q^N, \quad [a_i, a_j^\dagger] = 0 = [a_i, a_j], \tag{4}$$

which itself is not covariant under quantum group transformations. In addition, several inequivalent versions of Equation (4) are found in the literature. Some studies [16, 17] of $q$-bosons systems show that these type of deformations enhance BEC. However, the relevance of $q$-boson deformations, in contrast to Equations (2) and (3), lies outside the context of quantum group invariance. In the $q \to 1$ limit, the matrix $R \to \mathbf{1}$, and Equations (2) and (3) become the familiar boson algebra

$$\phi_i \phi_j^\dagger - \phi_j^\dagger \phi_i = \delta_{ij},$$
$$\phi_i \phi_j - \phi_j \phi_i = 0.$$

For a given $\kappa$ a normalized state with $n_1$ particles of species 1 and $n_2$ particles of species 2 is defined by

$$\frac{1}{\sqrt{\{n_1\}!\{n_2\}!}} \overline{\Phi}_2^{n_2} \overline{\Phi}_1^{n_1} |0\rangle, \tag{5}$$

where the $q$-numbers $\{n\} = \frac{1-q^{2n}}{1-q^2}$ and the $q$-factorials $\{n\}!$ are defined $\{n\}! = \{n\}\{n-1\}\{n-2\}...1$. For purposes of the following calculation, it is useful to express the operators $\Phi_j$ in terms of standard boson operators $\phi_i$. The corresponding representation has to be consistent with equations (2) and (3), and it is is given by

$$\Phi_2 = (\phi_2^\dagger)^{-1}\{N_2\} \tag{6}$$
$$\overline{\Phi}_2 = \phi_2^\dagger \tag{7}$$
$$\Phi_1 = (\phi_1^\dagger)^{-1}\{N_1\}q^{N_2} \tag{8}$$
$$\overline{\Phi}_1 = \phi_1^\dagger q^{N_2}, \tag{9}$$

leading to the Hamiltonian

$$\mathcal{H}_B = \sum_\kappa \varepsilon_\kappa \{\phi_{1,\kappa}^\dagger \phi_{1,\kappa} + \phi_{2,\kappa}^\dagger \phi_{2,\kappa}\}, \tag{10}$$

with the bracket $\{x\}$ as defined below Equation (5). After a simple algebraic manipulation the grand partition function is given by the Equation

$$\mathcal{Z}_B = \prod_\kappa \sum_{m=0}^\infty (m+1) e^{-\beta \varepsilon_\kappa \{m\}} z^m, \qquad (11)$$

where $z = e^{\beta \mu}$ is the fugacity.

## 3 One- and two-dimensional models

For $D = 1$ and $D = 2$ we write in the thermodynamic limit

$$\ln \mathcal{Z}_B = \ln \left(1 + \sum_{n=1}^\infty (n+1) z^n \right) + \frac{2\pi^s L^D}{h^D} \int_0^\infty p^{D-1} \ln \left(1 + \sum_{n=1}^\infty (n+1) e^{-\beta\{n\}\varepsilon} z^n \right) dp, \qquad (12)$$

where $s = 0$ ($s = 1$) for $D = 1$ ($D = 2$), and, as usual, the divergence of the ground state term at $z = 1$ has been taken into account by splitting the $\kappa$ summation into a single term plus an integration. Defining, for $D = 1$ and $D = 2$, the functions

$$F_D(z, q) = \int_0^\infty x^{D-1} \frac{\sum_{n=1}^\infty n(n+1) P_n(x, z)}{f(z, q)} dx, \qquad (13)$$

where we have defined $P_n(x, z) = e^{-\beta\{n\}x^2} z^n$ and

$$f(z, q) = 1 + \sum_1^\infty (m+1) e^{-\{m\}x^2} z^m,$$

the average total number of particles $\langle N \rangle = (1/\beta)(\partial \ln \mathcal{Z}_B / \partial \mu)_{T,V}$ is simply given by

$$\langle N \rangle = \langle N_0 \rangle + \frac{2 L^D}{\pi^{(1-s)/2} \lambda_T^D} F_D(z, q), \qquad (14)$$

with $\lambda_T = \sqrt{h^2/2\pi m k T}$. In the $q \to 1$ limit the bracket $\{n\} \to n$ and the functions $F_D(z, q)$ become the well known functions $g_m(z) = \sum_1^\infty z^k/k^m$ according to

$$F_1(z, 1) = \sqrt{\pi} g_{1/2}(z), \qquad (15)$$

$$F_2(z,1) = g_1(z), \tag{16}$$

and Equations (14) become those for an ideal Bose gas of two species in one and two dimensions. As it is well known, BEC is not possible for an ideal Bose gas in one and two dimensions because the functions $g_{1/2}(z)$ and $g_1(z)$ diverge as $z \to 1$. Since $f(1,q) > 1$, the convergence of the functions $F_D(1,q)$ can be checked by considering the behavior of the functions

$$\mathcal{F}_D(1,q) = \int_0^\infty x^{D-1} \sum_{n=1}^\infty n(n+1)e^{-\{n\}x^2} dx > F_D(1,q). \tag{17}$$

An elementary integration leads, for $q > 1$, to the convergent series

$$\mathcal{F}_D(1,q) = \frac{\Gamma(D/2)}{2} \sum_{n=1}^\infty \frac{n(n+1)}{\{n\}^{D/2}}. \tag{18}$$

Therefore, in contrast to the $q = 1$ case, the functions $F_D(1,q)$ converge for all values $q > 1$, showing then that the requirement of $SU_q(2)$ symmetry leads to BEC in one and two dimensions. The BEC of these systems will occur for those values of the temperature or density such that

$$\lambda^D \geq \frac{2F_D(1,q)}{\pi^{(1-s)/2}\langle n \rangle}, \quad D = 1, 2. \tag{19}$$

The critical temperature $T_c$ is found by setting $z = 1$ in Equation (14), and it is given by

$$T_c = \frac{h^2}{2\pi^s mk}\left(\frac{\langle n \rangle}{2F_D(1,q)}\right)^{2/D}. \tag{20}$$

Figure 1 shows a graph obtained from a numerical calculation of the function $t_c = \pi^{1-s}/F_D^{2/D}$ for $D = 1$ and $D = 2$. The critical temperature $T_c$ increases considerably for small values of $q > 1$ and become practically constant for $q \gg 1$.

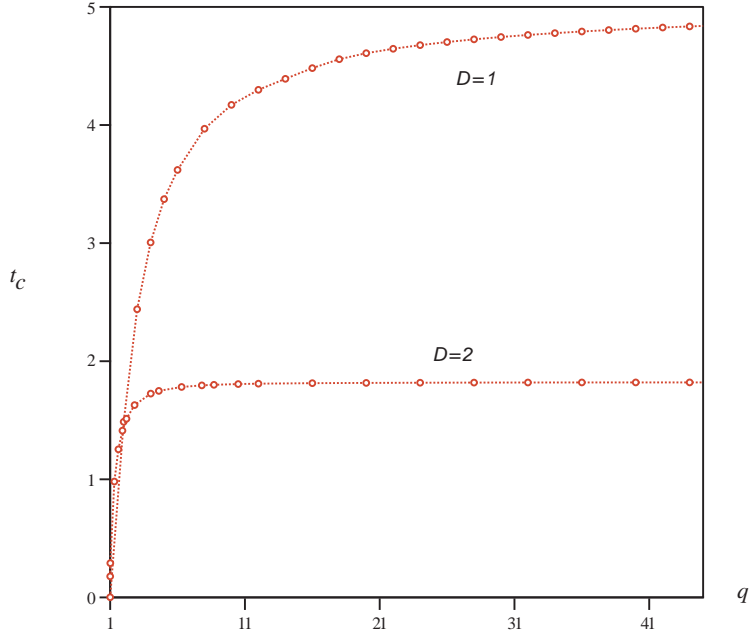

FIG. 1. The dependence of the function $t_c$, as defined in the text, on the parameter $q$ for one and two dimensional systems with $SU_q(2)$ symmetry. The behavior of the functions $F_D$ in Equation (20) is such that the critical temperature depends strongly for those values of $q$ closer to 1, and it becomes approximately constant for large values of $q$.

The internal energy $U$ is found from the equation

$$U = \frac{-\partial \ln \mathcal{Z}_B}{\partial \beta} + \mu \langle N \rangle$$
$$= \frac{2L^D}{\pi^{(1-s)/2} \lambda_T^D \beta} G_D(z, q), \quad (21)$$

where the functions $G_D(z, q)$ are defined according to

$$G_D(z, q) = \int_0^\infty x^{D+1} \frac{\sum_{n=1}^\infty (n+1)\{n\} P_n(x, z)}{f(z, q)} dx. \quad (22)$$

In the $q \to 1$ limit the functions $G_D$ become

$$G_1(z, 1) = \frac{\sqrt{\pi}}{2} g_{3/2}(z), \tag{23}$$

$$G_2(z, 1) = g_2(z). \tag{24}$$

In particular, the functions $G_1(1, q)$ and $G_2(1, q)$ decrease from their maximum values at $q = 1$; $(\sqrt{\pi}/2)\zeta(3/2) = 2.315$ and $\zeta(2) = \pi^2/6$ respectively, where $\zeta(z)$ is the Riemann zeta function.

For $q \neq 1$, the functions $F_D(z, q)$ and $G_D(z, q)$ are related each other in a similar way as the functions $g_n(z)$ and $g_{n-1}(z)$ do. From the simple identity

$$\int_0^\infty \frac{d}{dx}\left(x^D \frac{\sum_1^\infty (n+1)n P_n(x, z)}{f(z, q)}\right) dx = 0, \tag{25}$$

we find that for all values of $q > 1$

$$z \frac{d}{dz} G_D(z, q) = \frac{D}{2} F_D(z, q). \tag{26}$$

The behavior of the heat capacity $C_V$ is simply found by considering that the chemical potential $\mu(T)$ and its derivative $\mu' = \partial\mu/\partial T$ vanish for $T < T_c$. For $D = 1$ and $D = 2$, $C_V(T < T_c)$ is given by

$$C_V(T < T_c) = \frac{D+2}{2} \langle N \rangle k \left(\frac{T}{T_c}\right)^{D/2} \frac{G_D(1, q)}{F_D(1, q)}, \tag{27}$$

and for $T > T_c$

$$C_V(T > T_c) = \langle N \rangle k \left(\frac{D+2}{2} \frac{G_D(z, q)}{F_D(z, q)} - \frac{D^2}{4} \frac{F_D(z, q)}{H_D(z, q)}\right), \tag{28}$$

with the function $H_D(z, q)$ given by

$$H_D(z, q) = \int_0^\infty x^{D-1} \left(\frac{\sum_1^\infty (n+1)n^2 P_n(x, z)}{f(z, q)} - \frac{(\sum_1^\infty (n+1)n P_n(x, z))^2}{f^2(z, q)}\right) dx. \tag{29}$$

The gap in the heat capacity at the critical temperature is given by the second term in Equation (28) at $z = 1$. Figure 2 shows the gap $\Delta C_V/\langle N \rangle k$ dependence on $q$ produced by the discontinuity in the heat capacity at $T = T_c$ for $D = 1$ and $D = 2$. As previously discussed, large changes occur for those values near the standard value $q = 1$. For $q \gg 1$, the functions $F_D$, $G_D$ and $H_D$ become approximately constant. On the other hand, as $q \to 1$

$$\frac{F_D(z,q)}{H_D(z,q)} \to \frac{1}{z(d\ln F_D(z,1)/dz)}, \tag{30}$$

which vanishes at $z = 1$ and, in agreement with the ideal Bose case, no discontinuity appears.

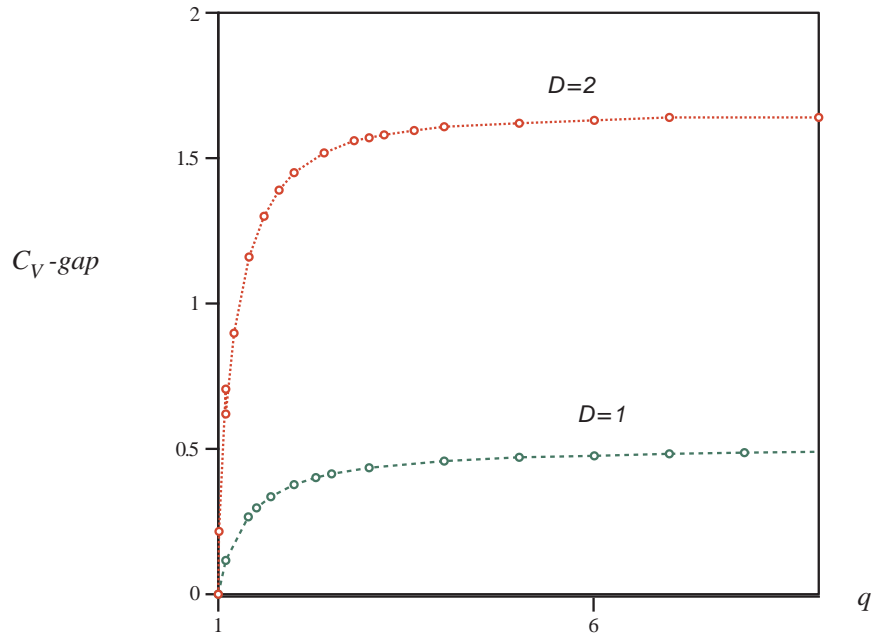

FIG. 2. The function $\Delta C_V/\langle N \rangle k$ showing the gap in the heat capacity as a function of $q$, that results from a numerical calculation of the second term in Equation (28) for $D = 1$ and $D = 2$.

The entropy function $S$ is calculated from

$$S = \frac{U - \mu \langle N \rangle}{T} + k \ln \mathcal{Z}_B. \qquad (31)$$

From Equations (12), (14), (21) and the identity

$$\int_0^\infty \frac{d}{dx} x^D \ln\left(1 + \sum_1^\infty (n+1) P_n(x,z)\right) dx = 0, \qquad (32)$$

we find for the entropy

$$S(T < T_c) = \frac{2kL^D \pi^{(s-1)/2}}{\lambda_T^D} G_D(1,q)\left(1 + \frac{2}{D}\right), \qquad (33)$$

and

$$S(T > T_c) = \frac{2kL^D \pi^{(s-1)/2}}{\lambda_T^D} \left[G_D(z,q)\left(1 + \frac{2}{D}\right) - F_D(z,q) \ln z\right]. \qquad (34)$$

Equation (34) reduces, for $q = 1$ and all values of $T \geq 0$, to the entropy of an ideal Bose gas with two species

$$S = \frac{kL^D}{\lambda_T^D}\left[(s+3)g_{\frac{D+2}{2}}(z) - 2g_{D/2}(z) \ln z\right] \qquad (35)$$

Figures 3 and 4 show the dependence of the functions $s_D = \frac{S}{k}(\frac{h}{L})^D(\frac{1}{2\pi mk})^{D/2}$ on the temperature $T < T_c$ for $q = 2$ and $q = 6$ in comparison to the entropy of an ideal Bose gas. The behavior of the functions $G_D(1,q)$ are such that, at low temperatures, the entropy function in Equation (33), lies below than the entropy of an ideal Bose gas, as given in Equation (35).

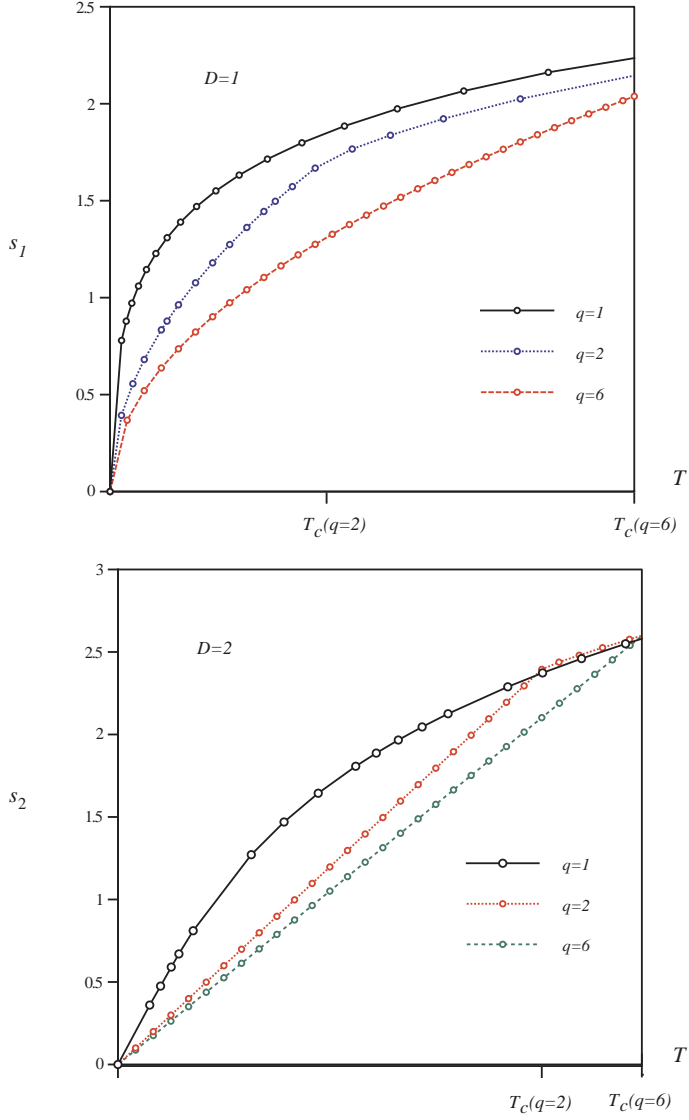

FIGS. 3,4. The entropy functions below the critical temperature for $q = 2$ and $q = 6$. For a given value of the parameter $q$, the entropy function lies below the Bose entropy for all temperatures $T < T_c(q)$.

One can use a simple analytical argument to compare Equation (33) with Equation (35) for low temperatures, $z \approx 1$. For example, for $D = 1$ and $z \approx 1$, we can approximate Equation (33) to the expression

$$S(T < T_c) \approx \frac{6kL}{\sqrt{\pi}\lambda_T}\left(G_1(z,q) - \frac{1}{2}F_1(1,q)(z-1)\right). \tag{36}$$

As previously discussed, since the functions $F_D(z,q)$ and $G_D(z,q)$ are more convergent when $q > 1$, we can write the following inequalities

$$F_1(z,1) > F_1(z,q), \tag{37}$$

and

$$G_1(z,1) > G_1(z,q). \tag{38}$$

By taking into account that $\ln z \approx z - 1$ for $z \approx 1$, and that independently of the value of $q$ the function $F_1(z,q) < F_1(1,q)$, and combining Equations (15), (23), (37) and (38) we obtain that

$$3g_{3/2}(z) > \frac{6}{\sqrt{\pi}}G_1(z,q), \tag{39}$$

and

$$2g_{1/2}(z)\ln z < \frac{3}{\sqrt{\pi}}F_1(1,q)(z-1), \tag{40}$$

which clearly imply that at low temperatures, the entropy for $q > 1$ is lower than the entropy of an ideal Bose gas. For $D = 2$, a similar analysis leads to the same conclusion as the $D = 1$ case.

Therefore, the requirement of $SU_q(2)$ symmetry has the additional consequence of reducing the entropy for all values of $q > 1$ and $T < T_c(q)$. This same behavior was previously found in Reference [14] for the case of a quantum group fermion gas, and in Reference [18] for the three dimensional case of the model in this paper.

## 4 Conclusions

The results reported in this paper show that, in the thermodynamic limit, the simplest models with $SU_q(2)$ invariance exhibit BEC in one and two dimensions. Since no interaction terms between the $SU_q(2)$ operators $\Phi_i$ appear in the Hamiltonian, as far as quantum group invariance is concerned, these models can be considered as 'free' models. However, in terms of boson operators, the Hamiltonian acquires interaction terms involving powers of boson number operators and $\ln q$. These interaction terms are certainly fixed by the quantum group, which are the values of $q$ and $N$ in $SU_q(N)$. Our results point out that the critical temperature $T_c$ and the gap in the heat capacity $C_V$ increase with the value of $q$, and that the thermodynamic properties of the model remain unchanged for very large values, $q > 40$. In addition, the $q$-dependence of the entropy function is such that it acquires its maximum value for the case of an ideal Bose gas, $q = 1$. In particular, for $D = 2$ this system has different behavior at high and low temperatures. While at low temperatures and $q > 1$ it exhibits BEC, it behaves as a fermion gas at high temperatures and $q > \sqrt{2}$ [15]. On the other hand, at low temperatures and $0 < q \leq 1$ no BEC is realized, and at high temperatures and $0 < q < \sqrt{2}$ it behaves as a boson gas. Therefore, the simple models studied here, show that the boson interaction terms induced by the requirement of quantum group invariance, lead to Bose-Einstein condensation without the necessity of including an external potential.